\documentclass[11pt]{article}

\usepackage[a4paper, top=3cm, bottom=3cm, left=2.5cm, right=2.5cm]{geometry}
\usepackage[english]{babel}
\usepackage[utf8]{inputenc}
\usepackage{graphicx}
\usepackage{amsthm}
\newtheorem{dfn}{Definition}
\newtheorem{theorem}{Theorem}
\usepackage{amsfonts}
\usepackage{amssymb}
\usepackage{amsmath}
\usepackage{multirow}
\usepackage{booktabs}
\usepackage{lscape}

\usepackage{tikz}
\definecolor{vtst}{RGB}{186,248,3}
\definecolor{vtnv}{RGB}{93,164,45}
\definecolor{trnv}{RGB}{151,226,177}
\definecolor{trt}{RGB}{189,219,236}
\definecolor{otqt}{RGB}{206,160,31}
\definecolor{otst}{RGB}{109,153,149}
\definecolor{otot}{RGB}{145,152,18}
\definecolor{otcq}{RGB}{214,58,48}
\definecolor{nvnt}{RGB}{189,9,184}
\definecolor{uno}{RGB}{237,158,178}
\definecolor{nove}{RGB}{190,225,85}
\definecolor{stuno}{RGB}{102,70,178}
\definecolor{qrt}{RGB}{41,50,54}

\newcommand{\mycbox}[1]{\tikz{\path[draw=#1,fill=#1] (0,0) rectangle (0.2cm,0.2cm);}}

\begin{document}
 
\title{Cluster analysis of weighted bipartite networks:\\
a new copula-based approach} 

\author{
Alessandro Chessa, Irene Crimaldi, Massimo Riccaboni, Luca Trapin
\footnote{alphabetic order} 
\footnote{IMT Institute for Advanced Studies Lucca, 
Piazza S. Ponziano, 6, 55100 Lucca, Italy}
\footnote{E-mail: alessandro.chessa@imtlucca.it,
  irene.crimaldi@imtlucca.it, massimo.riccaboni@imtlucca.it,
  luca.trapin@imtlucca.it} } 

\date{\today}

\maketitle

\begin{abstract}
In this work we are interested in identifying clusters of ``positional
equivalent'' actors, i.e. actors who play a similar role in a
system. In particular, we analyze weighted bipartite networks that
describes the relationships between actors on one side and features or
traits on the other, together with the intensity level to which actors
show their features. The main contribution of our work is
twofold. First, we develop a methodological approach that takes into
account the underlying multivariate dependence among groups of
actors. The idea is that positions in a network could be defined on
the basis of the similar intensity levels that the actors exhibit in
expressing some features, instead of just considering relationships
that actors hold with each others.  Second, we propose a new
clustering procedure that exploits the potentiality of copula
functions, a mathematical instrument for the modelization of the
stochastic dependence structure. Our clustering algorithm can be
applied both to binary and real-valued matrices. We validate it with
simulations and applications to real-world data.

\end{abstract}

\section{Introduction}
In the last few years complex network theory has attracted the
interest of a widespread audience as a powerful tool to analyze
complex relational structures and to represent big data
\cite{lohr12}.\\ \indent Network analysis usually deals with a massive
amount of data that requires to be managed and organized efficiently
in order to extract as much information as possible, reducing the
dimensionality of the problem. One of the most important methodologies
to tackle this issue is the identification of network communities
\cite{cerina2014}, \cite{fortunato10}, \cite{Girvan02},
\cite{newman01}.  {\em Community detection} allows us to extract
sub-networks which exhibit different properties from the aggregate
properties of the whole network and also to investigate information on
groups of nodes with similar cha\-rac\-te\-ri\-stics which are more
likely to be connected to each other. Communities are usually defined
as subsets of nodes that are densely connected, i.e., they are more
connected among themselves than to the rest of the network. However,
in many network applications, there is a meaningful group structure
which does not coincide with the partition into dense communities:
indeed, the groups may be characterized by similar patterns of
interactions with other groups \cite{Brandes07}. Within this context,
{\em positional analysis} is particularly interesting since it deals
with the identification of actors who occupy an equivalent position in
a system, i.e. play a similar role in the con\-si\-de\-red
organization. Differently to the community detection where the
clusters are represented by densely connected groups of actors,
positional analysis aims at studying relational data in order to
cluster the actors into some classes such that the elements of the
same class occupy equivalent positions in the system. In order to
illustrate the distinction between positional ana\-ly\-sis and community
detection, let us consider the following example of the e-mails sent
among the employees of a company: it may be that we are able to
identify different communities of individuals among which e-mails are
more frequently exchanged. However, densely connected employees may
occupy different positions in the organization and we need to run a
positional analysis if we are interested in identifying groups of
actors with equivalent positions.\\ \indent In this work we aim at
identifying clusters of ``positional equivalent'' actors in cases
where the available data are the relationships defined among actors on
one side and some features on the other one \cite{borgatti97},
\cite{borgatti08}, \cite{Brusco11}, instead of interpersonal
relationships. Basically, the idea is that positions in a network
structure can be defined according to the characteristics or
behaviours that the actors exhibit, instead of the relationships that
actors hold with other actors. Individual to attribute relations can
be represented as a {\em weighted bipartite network} where the
edge-weights represent the level to which actors show a particular
feature. More precisely, a network is bipartite if its nodes can be
divided into two sets in such a way that every edge connects a node in
one set to a node in the other one \cite{asratian98}. Bipartite
networks are thus very useful for representing data in which the
elements under scrutiny belong to two categories (typically referred
to as actors, or agents, and features, respectively), and we want to
understand how the elements in one category are associated with those
in the other one. Notable examples that have been analyzed include
networks of company directors and the board of directors on which they
sit \cite{Davis97}, \cite{Mariolis75}, scientific collaboration
networks \cite{barabasi02}, \cite{Guimera05}, \cite{newman01},
networks of documents and words \cite{dhillon01}, as well as network
of genes and genetic sequences \cite{larremore13}.\\ \indent The
widespread approach to partition bipartite networks consists of
applying standard community detection algorithms, such as the
Girvan-Newman modularity \cite{Girvan02}, to the one-mode projection
of the original network. Consider two types of nodes, say {\it a} and
{\it b}, in a one-mode projection of the bipartite network, nodes of
the same type, say {\it a}, are connected to each other if they share
a common node of the other type, say {\it b}. For instance, in the CEO
network, two CEOs are connected if they both sit in the same
board. Although the one-mode projection procedure can give some
insights on the topological properties of the network, at the same
time it can imply the lost of relevant information. In fact, different
bipartite networks may reduce to the same one-mode projection, and
thus a clustering based on the latter may produce unreliable or
incorrect results, as shown in \cite{Good10} and \cite{Zhou07}.
Regardless of those critiques, in \cite{everett13} the authors argue
that under some circumstances, using multiple projections, the
information extracted with this procedure is sound, and therefore the
simplicity of this approach can be still exploited. However, several
authors tried to solve this problem by defining measures and
algorithms that could be directly applied to the original matrix
associated to the bipartite network.\\ \indent In the physics
community, two different definitions of bipartite modularity have been
proposed, \cite{barber07}, \cite{Guimera07}. Both concepts extend the
Girvan-Newman modularity, but pose different assumptions on the null
model taken as the benchmark in the metric used for the module
identification. They return good results compared to the one-mode
projection, but their applicability is restricted to the case of
binary bipartite networks. \\ \indent Some applications of bipartite
networks refer to affiliation networks \cite{borgatti11}, which
capture social relationships, such as membership or event
participation. Positional analysis is well established in social
network literature, where the usual approach consists of applying the
standard measures of structural or regular equivalence, and the
related algorithms, to the one mode-projection of the affiliation
network \cite{wasser94}. However, affiliation networks represent only
a very special case of bipartite networks since the associated
matrices are binary.\\ \indent Other proposed methods for bipartite
network clustering, that are mostly used by sociologists, are based on
blockmodeling (e.g. \cite{Brusco06}, \cite{Brusco07}, \cite{doreian04}
and \cite{zib07}).  The key idea of this approach is that the rows and
the columns of the matrix associated to the bipartite network can be
partitioned simultaneously by means of a criterion function, which
measures the inconsistencies of the empirical blocks with the ideal
ones. Therefore, blockmodeling works directly on the matrix by trying
to permute rows and columns in order to fit, as closely as possible,
idealized pictures. The differences between the various types of
blockmodeling techniques concern the definition of the ideal blocks
and the criterion functions. Blockmodeling is mostly applied to binary
data, but it can also be exploited for weighted matrices (valued
blockmodeling and homogeneity blockmodeling \cite{zib07}). However,
with the valued blockmodeling, information about the values above a
pre-specified parameter is lost and a problem is to determine
appropriately the value of this parameter. The homogeneity
blockmodeling does not require any additional parameters to be set in
advance and it uses all available information, but its main
disadvantage is that it can consider only a few possible ideal
blocks.\\ \indent In \cite{larremore14}, the authors proposed a
bipartite stochastic block model where a parametric
pro\-ba\-bi\-li\-stic structure is given, and the clusters are
identified by solving the inference problem of finding the parameters
that best fit the observed network. In particular, they model the
generating process of the number of edges between two nodes of
different types with a Poisson distribution with a certain intensity
parameter. The authors show that their method outperforms the one-mode
projection approach. Nevertheless, it does not deal with the case when
we have weights on the edges. In \cite{aicher14}, the authors try to
go in this direction by proposing a stochastic block model for
edge-weighted networks, but their method requires to choose the number
of clusters (as in most stochastic block models).  \\ \indent The
algorithm we propose realizes a partition of ``positional equivalent''
actors based on the entire information enclosed in the weighted
bipartite network that describes their cha\-rac\-te\-ri\-stics or
behaviours. The main contribution of our work is twofold. First, we
develop a new methodological approach according to which actors are
grouped with respect to their intrinsic multivariate stochastic
dependence structure.  In this framework, not only the magnitude of a
single weight matters but the whole pattern of the values the actors
show along all the features is relevant for the
classification. Se\-cond, we propose a new clustering procedure that
exploits the potentiality of copula functions, a mathematical
instrument for the modelization of the multivariate stochastic
dependence structure. In particular, copulas allow us to group actors
according to their underlying dependence structure, without any
assumption on their one-dimensional marginal distributions, and to
take into account various kinds of stochastic dependence structures
among actors. Moreover, there is no need to predefine the target
number of clusters.\\ \indent The paper is structured as follows. In
Sections \ref{our} and \ref{procedure}, we describe our approach,
together with the mathematical tool we employ, and then we illustrate
our clustering procedure. In Sections \ref{simul} and \ref{data} we
show the performance of our clustering algorithm applying it to
simulated and real data. Finally, in Section \ref{fine} we conclude
with a discussion of the potentiality of our method and possible
future applications and extensions.

\section{A copula-based approach}
\label{our}
As explained in the previous section, we consider the general setting
where we have an $N \times M$ real-valued matrix, that collects the
information on the connections that go from a set of $N$ actors to a
set of $M$ items, representing some features or behaviours. The
elements of such a matrix can be any real numbers, with zero
representing the absence of a relationship and a non-zero value
representing the presence of a relationships, together with its
intensity. As an example, this framework can be used to analyse
situations where we have actors on one side and personal qualities or
interests on the other side, and the weighted-edges between the two
sets can be used to represents the level to which an individual shows
a certain quality or interest. Another example may be a set of
individuals in a supermarket and the set of products they buy. In this
case, an edge represent whether an individual bought a particular
product or not, and its value gives the amount of product bought or
its cost.\\ \indent Against this background, we want to emphasize that
actors may be classified into positions based on their patterns of
characteristics, interests or behaviours that they exhibit and on the
intensity wherewith the actors show them, instead of the kind of
relationships that they keep with other actors. In other words, we
move in the direction that the dependence (we mean positive
dependence, i.e. similarity) in the expression levels of the
considered features is related to the position that the actors occupy
in the system. Hence, we say that some actors are {\em positional
  equivalent} if they show a significant dependence structure that
join them. In this framework, the use of the traditional one-mode
projection methods would be meaningless and misleading and also
blockmodeling or modularity approaches adapted to bipartite networks
could not give a clear answer to the problem because they are not well
tailor made for the analysis of weighted bipartite networks.\\ \indent
Our purpose is to identify clusters of actors by means of the
detection, from the original matrix, of some statistically significant
dependencies among groups of actors. Basically, our assumption is that
actors within a system have an underlying multivariate stochastic
dependence structure which generates the data.  In order to identify
this intrinsic dependence structure, we propose to exploit the
mathematical copula theory.\\ \indent The concept of copula was
introduced during the forties and the fifties with Hoeffding
\cite{hoeffding40} and Sklar \cite{sklar59}, but the evidence of a
growing interest in this kind of functions in statistics started only
in the nineties \cite{nelsen06}. Copulas are functions that join or
``couple'' multivariate distribution functions to their
one-dimensional marginal distributions. More precisely, we have the
following definition and results\footnote{For more details, we refer
  to the various excellent monographs existing in literature, such as
  \cite{joe97}, \cite{nelsen06} and \cite{trivedi}.}.
\begin{dfn}
A $d$-dimensional copula $C({\bf u})=C(u_1,...,u_d)$ is a function
defined on $[0,1]^d$ with values in $[0,1]$, which satisfies the
following three properties:
\begin{enumerate}
\item $C(1,...1,u_i,1,...,1)=u_i$ for every $i \in \{1,...,d\}$ and $u_i
  \in [0,1]$;
\item if $u_i=0$ for at least one $i$, then $C(u_1,\dots,u_d)=0$;
\item for every $(a_1,...,a_d),(b_1,....b_d) \in [0,1]^d$ with   
$a_i\leq b_i$ for all $i$, 
\[ 
\sum^2_{j_1=1}...\sum^2_{j_d=1}(-1)^{j_1+...+j_d}C(u_{1,j_1},...,u_{d,j_d}) \geq 0
\]
where, for each $i$,  $u_{i,1}=a_i$ and $u_{i,2}=b_i$.
\end{enumerate}
\end{dfn}
The advantage of the copula functions and the reason why they are used
in the dependence modeling is related to the Sklar's theorem
\cite{sklar59}. It essentially states that every multivariate
cumulative distribution function can be rewritten in terms of the
margins, i.e. the marginal cumulative distribution functions, and a
copula.
\begin{theorem}
Let $F$ be a multivariate cumulative distribution function with
margins $F_1,...,F_d$. Then there exists a copula $C:[0,1]^d
\rightarrow [0,1]$ such that, for every 
$x_1,...,x_d\in {\overline {\mathbb{R}}}=[-\infty,+\infty]$, we have
\begin{equation}\label{sklar-eq}
F(x_1,...,x_d)=C(F_1(x_1),...,F_d(x_d))
\end{equation}
If the margins $F_1,\dots, F_d$ are all continuous, then $C$ is
unique; otherwise $C$ is uniquely determined on
$F_1({\overline {\mathbb{R}}})\times \cdots\times
F_d({\overline  {\mathbb{R}}})$.\\ Conversely, if $C$ is a copula and
$F_1,\dots,F_d$ are cumulative distribution functions, then $F$
defined by (\ref{sklar-eq}) is a multivariate cumulative distribution
function with margins $F_1,\dots,F_d$.
\end{theorem}
In the case when $f$ and $f_1,\dots, f_d$ are the marginal probability
density functions associated to $F$ and $F_1,\dots, F_d$,
respectively, the copula density $c$ satisfies
\begin{equation*}
f(x_1 ,\dots, x_d)=c\big(F_1(x_1),\dots, F_d(x_d)\big)\prod_{i=1}^df_i(x_i).
\end{equation*}
There are different families of copula functions that capture
different aspects of the dependence structure: positive and negative
dependence, symmetry, heaviness of tail dependence and so on. In our
work, we limit ourselves to the principal copula functions of the
Archimedean fa\-mi\-ly (namely, Gumbel, Clayton and Frank copulas, see
Appendix \ref{copula} for their definitions), which model, through a
unique parameter $\theta$, situations with different degrees of
dependence. Nonetheless, it is worth to note that the application of
our methodology is not restricted to those copula functions.

\section{Methodology}
\label{procedure}

In this section we present a copula-based technique that realizes a
partition of actors into clusters so that the actors belonging to the
same cluster show a significant dependence structure that allows us to
classify them as being ``positional equivalent''. Our approach is
inspired by the work of Di Lascio and Giannerini \cite{diLascio12a},
which introduced and studied a copula-based clustering algorithm,
called CoClust, in the framework of microarray data in genetics. As
they did, we use copula functions in order to model the multivariate
stochastic dependence structure among groups of actors and we apply
the maximized log-likelihood function criterion for the detection of
the different clusters.  Notwithstanding, our algorithm presents the
following important differences with respect to the one proposed by Di
Lascio and Giannerini:
\begin{itemize}
\item[1)] while they assume independence within clusters and
  dependence between clusters, we look for clusters
  of dependent actors;
\item[2)] while they first find the optimal number $K$ of clusters and
  then perform sequential extractions of $K$ actors, where at each
  time one actor is added to each cluster in a certain way, we do not
  use a sequential extraction method but we directly look for the
  optimal partition of the actors into clusters;
\item[3)] differently from them, we allow clusters to be of different
  sizes and we allocate all the actors into the clusters;
\item[4)] whereas they assume identity in distribution for actors
  inside a certain cluster, i.e. each cluster identifies one margin,
  we do not make this assumption and we estimate for each actor his
  own cumulative distribution function.
\end{itemize}
Given $N$ actors and $M$ items, we can represent the data that
describe the relationships between actors and items with a real-valued
matrix of dimension $ N \times M $,
\begin{equation*}
\begin{bmatrix} 
x_{11} & \cdots & x_{1m} & \cdots & x_{1M} \\ 
\vdots & \ddots & \vdots & \ddots & \vdots\\ 
x_{i1} & \cdots & x_{im} & \cdots & x_{iM}\\
\vdots & \ddots & \vdots & \ddots & \vdots\\ 
x_{N1} & \cdots & x_{Nm} & \cdots & x_{NM}
\end{bmatrix}
\end{equation*}
where $x_{im}$ represents the value of the item $m$ for the actor
$i$. With the language of network theory, this matrix can be seen as
the matrix associated to a weighted bipartite network.\\

\indent The procedure we propose takes as input this matrix and
returns the optimal decomposition into clusters after the following
four steps:
\begin{enumerate}
\item First of all, we find the empirical cumulative distribution
  function 
$$
\widehat F_i(x)=\frac{1}{M}\sum_{m=1}^MI_{\{x_{im}\leq x\}}
$$ of every actor $i$ based on the corresponding $M$-dimensional row
$x_{i\cdot}$ of the items.  For each actor $i$, we are taking the
values $x_{i1},\dots, x_{iM}$ of the $M$ items as i.i.d. realizations
drawn from the same univariate distribution. \\

\item The second step consists in the estimation of the copula
  objective function for each possible group $\mathcal C$ of actors,
  with $card({\mathcal C})\geq 2$, using pseudo-maximum likelihood
  estimation in order to estimate the dependence parameter. Thus, for
  each possible group, say ${\mathcal C}=\{i_1,\dots,i_k\}$, with
  $2\leq k\leq N$, of actors, we maximize the log-likelihood function
  defined as
\begin{equation*}
\theta\mapsto\ell_{\mathcal C}(\theta)=\sum_{m=1}^M
\ln c\left(
\widehat F_{i_1}(x_{i_1m}),\dots, \widehat F_{i_k}(x_{i_km}); \theta 
\right)\,,
\end{equation*}
where $c(u_1,\dots, u_k;\theta)$ denotes the parametric expression for
the chosen copula density, and we find the value $\ell^*({\mathcal
  C})$ such that
\begin{equation*}
\ell^*({\mathcal C})=
\ell_{\mathcal C}(\widehat\theta)=\max_{\theta\in\Theta}\sum_{m=1}^M
\ln c\left(
\widehat F_{i_1}(x_{i_1m}),\dots, \widehat F_{i_k}(x_{i_km}); \theta 
\right)\,.
\end{equation*}
Note that we are taking the vectors $\{(x_{i_1m},\dots, x_{i_{k}m}):
m=1,\dots,M\}$ as $M$ i.i.d. realizations drawn from the same
$k$-variate distribution.

\item In the third step, we consider the set $\mathcal P$ of all
  possible partitions of the $N$ actors that do not contain clusters
  with a single actor. Hence, each $\pi$ in $\mathcal P$ is formed by
  a certain number of clusters ${\mathcal C}$ with $card({\mathcal
    C})\geq 2$. The set $\mathcal P$ represents the set of all
  possible decompositions into clusters that the procedure can
  return\footnote{For example, if we have $4$ actors, numbered from
    $1$ to $4$, the set $\mathcal P$ is formed by the following
    partitions: $\pi_1=\{{\mathcal C}_{1,1}=\{1,2\},{\mathcal
      C}_{1,2}=\{1,3\}\}$, $\pi_2=\{{\mathcal
      C}_{2,1}=\{1,3\},{\mathcal C}_{2,2}=\{2,4\}\}$,
    $\pi_3=\{{\mathcal C}_{3,1}=\{1,4\},{\mathcal C}_{3,2}=\{2,3\}\}$
    and $\pi_4=\{{\mathcal C}_{4,1}=\{1,2,3,4\}\}$.}.  Namely,
  we find the maximum value of the map defined on $\mathcal P$ by
\begin{equation*}
\mathcal{L}(\pi)=\sum_{{\mathcal C}\in \pi} \ell^*({\mathcal C}).
\end{equation*}

\item Finally, the procedure returns $\pi^*\in{\mathcal P}$ such that
\begin{equation*}
\mathcal{L}^*=\mathcal{L}(\pi^*)=
\max_{\pi\in{\mathcal P}}\mathcal{L}(\pi).
\end{equation*}
More precisely, it returns the clusters that form $\pi^*$ in a 
decreasing order with respect to the value $\ell^*({\mathcal
  C})$ of each cluster $\mathcal C$ in $\pi^*$.
\end{enumerate}

The R code for this procedure is available at 
http://riccaboni.weebly.com/code.html.

\section{Simulations} 
\label{simul}
In order to verify the accuracy of the proposed algorithm, we
conducted some simulation ex\-pe\-ri\-ments.  We considered different
scenarios and, for each of them, various values of $M$ (i.e. the
number of items): $M=20,\,50,\,100$ or $250$. For each scenario and
each value of $M$, we generated $50$ random samples of $N=10$
observations (actors), grouped into $3$ clusters. Specifically, we
simulated the following scenarios:
\begin{enumerate}
\item In the first three scenarios, clusters were generated
  using standard Gaussian margins and the same copula type: one scenario 
  with Gumbel, another with Clayton and another one with Frank. Two of
  the clusters had $3$ observations (actors) and the dependence
  parameter $\theta=4$. The last one had $4$ observations
  (actors) with dependence parameter $\theta=3$.

\item In the second three scenarios, we generate the clusters
  using discrete marginal distribution, namely the Poisson
  distribution with parameter $\lambda=4$, and the same copula type:
  one scenario with Gumbel, another with Clayton and another one with
  Frank. Two clusters had $3$ observations (actors) and
  dependence parameter $\theta=4$, while the third one had $4$
  observations (actors) with dependence parameter $\theta=3$.

\item In the last three scenarios, we decided to draw the observations
  from different margins and using the same copula type: one scenario
  with Gumbel, another with Clayton and another one with Frank. In
  particular, the first two clusters had dependence parameter
  $\theta=4$ and respectively, Pareto(1,2) and Lognormal(0,1) marginal
  distributions. Instead, in the third cluster the dependence
  parameter was $\theta=3$ with Exponential(0.5) marginal
  distribution.
\end{enumerate}

For each considered scenario and each value of $M$, we tested the
algorithm applying the same copula used for the simulation of the
data. Moreover, for the first three scenarios and each value of $M$,
we also applied the algorithm with the other two copula types than the
one used for simulations.\\ \indent We observed that the choice of the
copula for the algorithm has no great effect on the performance of the
algorithm and the results seem quite good, especially in the case when
$M=100$ or $M=250$. Some main remarks can be made:
\begin{itemize}
\item First of all, under all the possible scenarios, for $M=100$ or
  $M=250$, we always got a $100\%$ percentage of successes in
  recognizing the clusters correctly.

\item Second, when the observations are drawn from the Gumbel and the
  Clayton copulas, we got a percentage of successes equal to $100\%$
  already for $M=50$ and between $80\%$ and $100\%$ for $M=20$.

\item Finally, when the observations are drawn from the Frank copula,
  we notice some problems for $M=20$. Indeed, for this copula type,
  $20$ realizations are too few to generate an evident dependence
  structure and so the algorithm does not work well in recognized
  it. However, we observed a fast improvement for $M$ getting larger
  and, starting from $M=50$, we can say that the percentage of
  successes are good ($75-80\%$).
\end{itemize}

\section{Applications to real-world data}
\label{data}

In this section, we describe two applications of our algorithm to real
datasets. The first one deals with a real-valued bipartite
network. The second one refers to a widely studied social network that
is described by a signed network.

\subsection{Trade data}
\begin{figure}[!h]
\centering
\caption{Trade share plot} \includegraphics[scale=0.75]{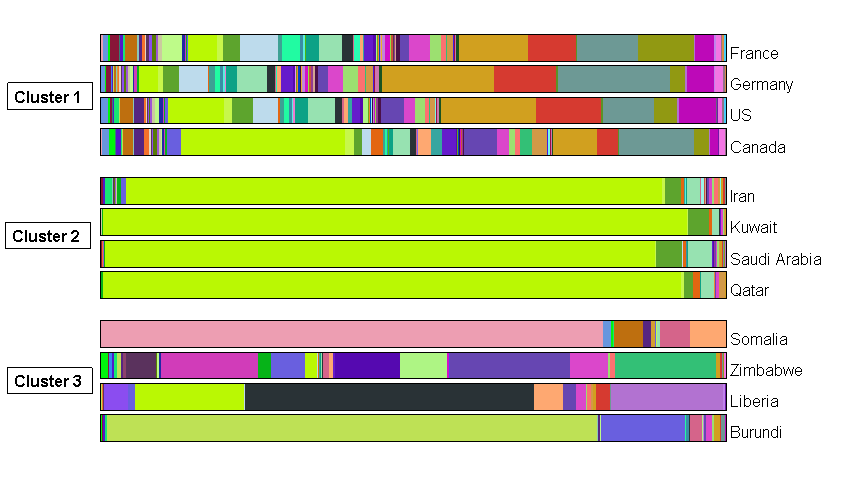}
        {\footnotesize In this figure we report for each country,
          classified in the relative cluster, a coloured bar
          representing the share of export for each of the $97$ HS2
          product categories over the total amount of
          export. Regarding {\it Cluster 1}, the high number of
          colours into the bar makes it clear that these countries use
          to trade in several product categories. Furthermore, an
          explicit dependence pattern arise from the proportion of the
          colours into the bars. In particular, the following product
          categories contributes to this strength relationship: {\bf
            84} (\mycbox{otqt}) - Nuclear reactors, Boilers, Machinery
          and mechanicals appliances; {\bf 87} (\mycbox{otst}) -
          Vehicles; {\bf 88} (\mycbox{otot}) - Aircraft and
          Spacecraft; {\bf 85} (\mycbox{otcq}) - Electrical machinery,
          Telecommunications equipment, Sound and Television
          recorders; {\bf 30} (\mycbox{trt}) - Pharmaceutical
          products; {\bf 90} (\mycbox{nvnt}) - Optical, Photographic,
          Cinematographic, Measuring, Checking, Precision, Medical
          instruments. Regarding {\it Cluster 2} the dependence
          relationship mainly arises from these three categories: {\bf
            27} (\mycbox{vtst}) - Mineral, Fuels, Oils; {\bf 29}
          (\mycbox{vtnv}) - Organic chemicals; {\bf 39}
          (\mycbox{trnv}) - Plastic and Articles thereof. However, it
          is important to remark that our clusering approach takes in
          consideration also the fact that these countries trade in a
          very small number of products, as can be seen from the few
          colours in the respective bars. The same reasoning apply for
          {\it Cluster 3} where, although the countries are
          specialized in a unique product such as category {\bf 9}
          (\mycbox{nove}) - Coffee, Tea, Mate and Spices for Burundi
          or category {\bf 1} (\mycbox{uno}) - Live animals for
          Somalia, the common pattern that makes them similar is the
          fact that they do not trade in most of the $97$ HS2
          categories.  }
\label{figure:trade}
\end{figure}

\begin{figure}[!h]
\centering
\caption{Trade network structure}
\includegraphics[scale=0.45]{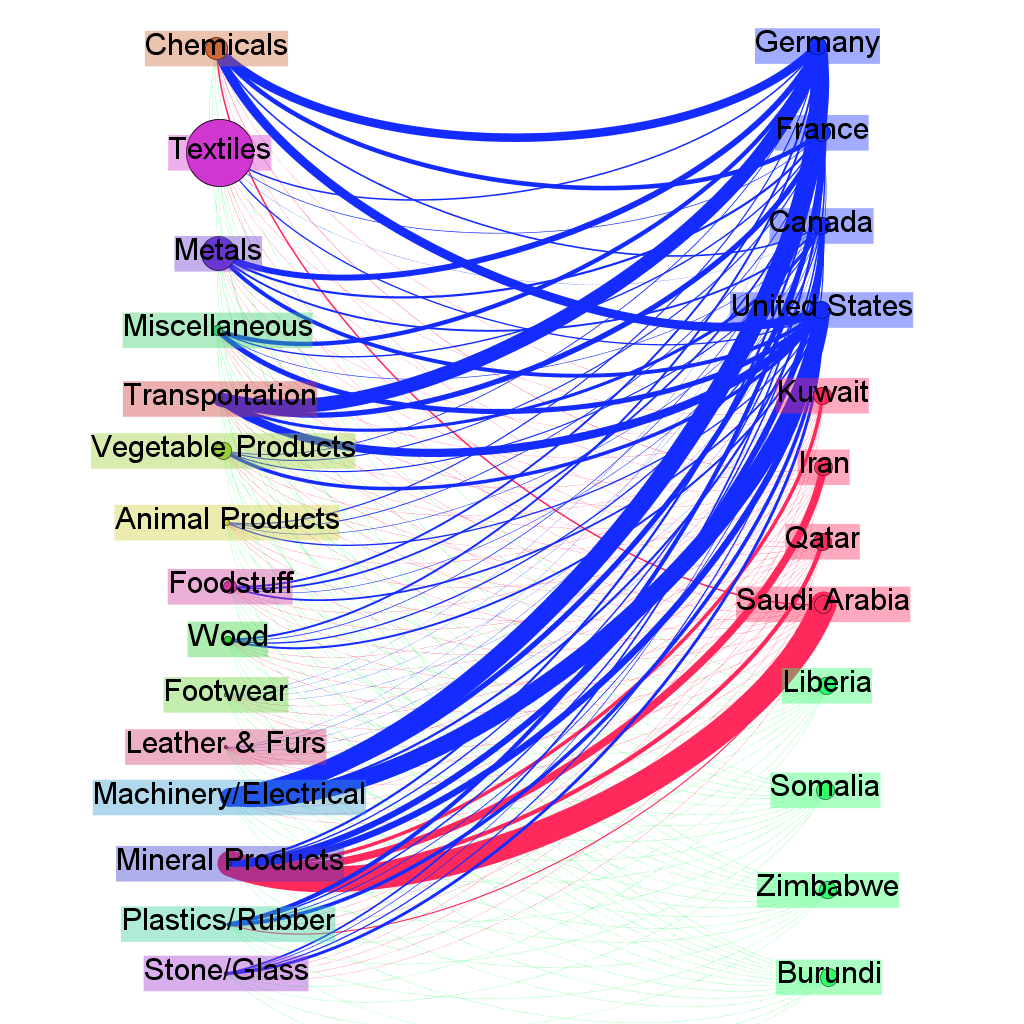}
\footnotesize{In this figure we show the weighted bipartite Trade
  network. On the right the three groups of countries, detected by our
  algorithm, and on the left the $97$ products categories, grouped in $15$ 
  homogeneus macro categories in order to highlight the relevant
  connections among the two different type of nodes. The size of the
  macro categories are in proportion to the number of categories
  grouped in them.  It is clear from the links partition how our
  metodology is able to disentangle different country categories
  according to the trade patterns, even for the third world countries
  (green background) for which the link weights are much smaller than
  the others.}
\label{figure:TradeNet}
\end{figure}

The first application we show is based on the BACI\footnote{French
  acronym of ``Base pour l'Analyse du Commerce International'':
  Database for International Trade
  Analysis.}-COMTRADE\footnote{Commodities Trade Statistics Database.}
dataset, featuring the amounts of import-export trades among several
countries in the world. We extracted a {\em weighted bipartite
  network} taking the export dollar values for the $M=97$ product
categories of the HS2 \footnote{The Harmonized System (HS) is an
  international nomenclature for the classification of products. It
  allows participating countries to classify traded goods on a common
  basis for customs purposes. It is a six-digit code system but we
  exploit the first two-digit in the analysis.} classification, for
selected $N=12$ countries, in the year $2011$. More in details, we
decided to select the countries according to their economies, in order
to identify $3$ hypothetical categories:
\begin{itemize}
\item a {\it First world} category composed by France, Germany, Canada
  and United states;
\item a {\it Third world} category represented by Burundi, Zimbabwe,
  Liberia and Somalia;
\item an {\it OPEC representative} category made by Kuwait, Saudi
  Arabia, Qatar and Iran.
\end{itemize}
We applied our procedure to the matrix, where the countries were in
rows, the products in columns, and each cell contained the gross
export value of a given country for a given product. Our aim was to
create clusters of countries which are similar (i.e. positional
equivalent in the International Trade Network) with respect to the
products they export. Much of the literature that focuses on
international trade looks for community detection, that is for
communities of countries with a high number of connections among them,
while being relatively less interconnected with countries outside the
community they are part of \cite{tzekina08}. Differently from the
classical clustering analysis in international trade, we tried to
define ``positional equivalent'' countries based on the products they
trade and not on the basis of the countries wherewith they trade.
Indeed, we were not interested in finding dense communities of
countries for different commodities, but we wanted to identify
countries that cover the same position in the trade network since they
present a similarity in their exports.\\ \indent The result we
obtained is reported in Table \ref{table:trade}. As we can see, the
algorithm is able to perfectly recognize the above mentioned country
groups. However, since these groups were built according to a
subjective judgement, we decided to analyze the data in order to
provide a more robust explanation for the clusters we found. In table
\ref{table:selection} we report for each country, the percentage on
the total amount of export\footnote{It is important to remark that,
  while in this table we report the percentage amounts for some
  selected product categories, we applied our algorithm directly on
  the export values for the all $97$ categories.}  for a selection of
$21$ HS2 categories out of the $97$ available, in order to give some
hints on the trade joint patterns that our algorithm recognize.
Overall, we can agree on the fact that the result is coherent with the
observed data.  Regarding the {\it First world} category, we can see
that at least a small amount of their total exports is allocated in
each selected categories and about the $60\%$ of their total export is
concentrated in the nine categories, corresponding to the following
commodities: {\bf 84} - Nuclear reactors, Boilers, Machinery and
mechanicals appliances; {\bf 87} - Vehicles; {\bf 88} - Aircraft and
Spacecraft; {\bf 85} - Electrical machinery, Telecommunications
equipment, Sound and Television recorders; {\bf 30} - Pharmaceutical
products; {\bf 90} - Optical, Photographic, Cinematographic,
Measuring, Checking, Precision, Medical instruments.\\ Conversely, for
the {\it OPEC Representative} group, it is clear that the nature of
the dependence arises from the fact that more than the $90\%$ of the
total export of these countries belongs to the following three
categories: {\bf 27} - Mineral, Fuels, Oils; {\bf 29} - Organic
chemicals; {\bf 39} - Plastic and Articles thereof. Nonetheless, we
underline that our algorithm did not recognize this cluster just
because of the large share of export these countries have in these few
products, but it captured the whole dependence between these countries
and so also the categories in which they do not trade, or trade a
little, play an important role. This is clear by looking at the
network structure for the {\it Third world} category in the last four
columns of table \ref{table:selection}. As it can be seen, all these
countries present a huge amount of the total export in a few specific
commodities. For example, more than the $80\%$ of the somalian export
is in category {\bf 1} - Live animals, while the $78\%$ of the
burundian export is in category {\bf 9} - Coffee, Tea, Mate and
Spices. Thus, we can affirm that these countries present a highly
specific production and the dependence among them arise not as a
consequence of the products in which they trade but rather from the
products in which they do not trade. By looking carefully at table
\ref{table:selection}, it is possible to notice that for most of the
selected 21 HS2 categories, the share of export is almost zero in all
these {\it Third World} countries. In this sense, they are similar to
the {\it OPEC representative} countries but, as we already said, the
latter present a specific dependence deriving from the common
commodities they trade. Finally, {\it Canada} deserves some
comments. It has an high value in category {\bf 27} as the countries
in the {\it Opec representative} category, but its values for the
other categories are more similar to those of the {\it First world}
than the ones of the {\it Opec representative} group. Our algorithm is
able to capture this aspect.\\ An insight of all these distinguishing
features between the clusters can also be grasped looking at figure
(\ref{figure:trade}), where we report for each country a coloured bar
with the export shares for each of the 97 HS2 product categories over
the total amount of export, and figure (\ref{figure:TradeNet}), where
we depict the bipartite trade network between the countries and 15
macro-categories of the HS2 products classification.

\begin{table}[h]
\caption{Trade data}
\centering
\begin{tabular}{c| cccc}
\toprule
{\bf Cluster 1} & France & Germany & United States & Canada \\ 
\midrule
{\bf Cluster 2} & Iran & Kuwait & Saudi Arabia & Qatar \\
\midrule
{\bf Cluster 3} & Burundi & Somalia & Zimbabwe & Liberia \\
\bottomrule
\end{tabular}
\label{table:trade}
\end{table}


\subsection{Supreme Court voting data}

\begin{figure}[!h]
\caption{Justice sentences network}
\includegraphics[scale=0.45]{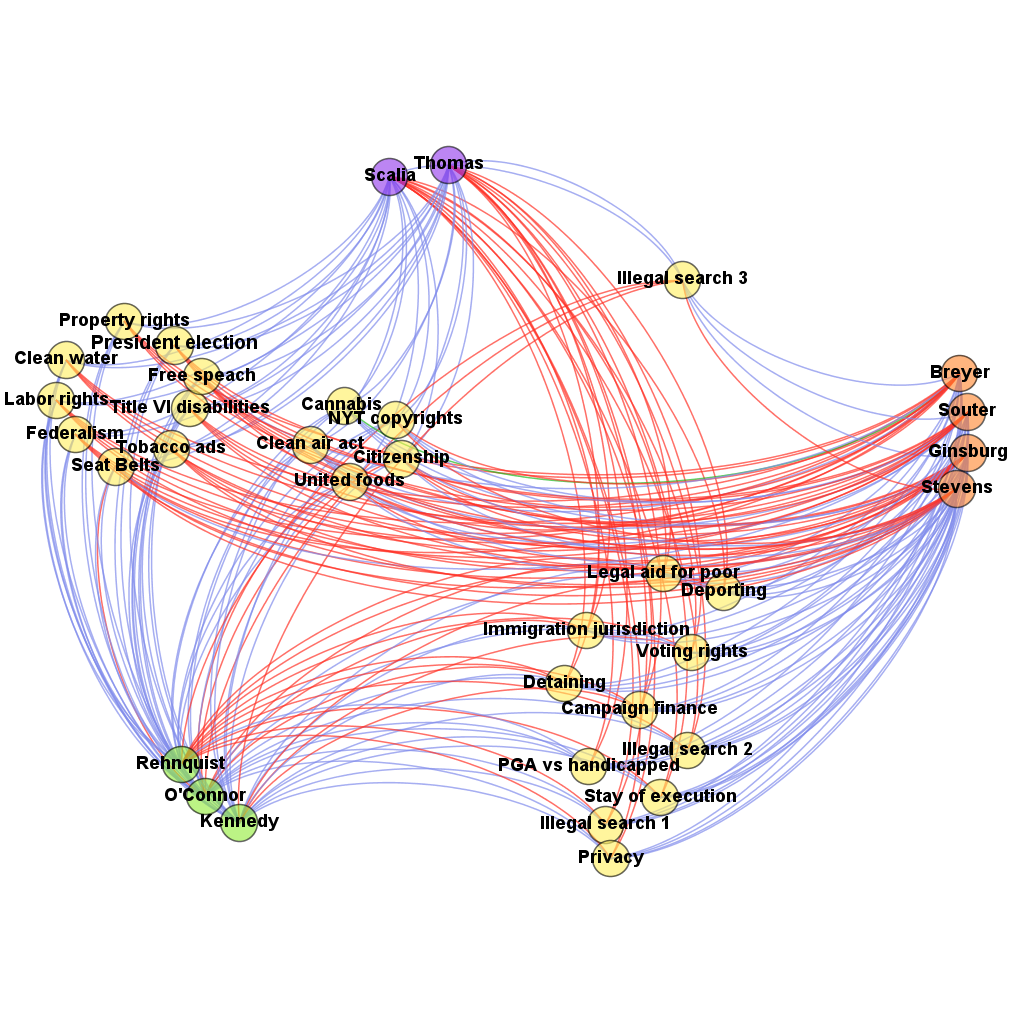} \centering {\footnotesize
  This figure depicts the bipartite signed network of the US Supreme
  Court Justice votes upon $26$ different issues. The blue edges
  correspond to votes in the majority ($+1$), the red edges correspond
  to votes in the minority ($-1$) and the unique green edge correspond
  to a case of abstention ($0$). Furthermore, the nodes are classified
  as follow: yellow for the $26$ issues, violet for {\it Cluster 1},
  orange for {\it Cluster 2} and green for {\it Cluster 3}. The
  network has been built so as to capture the sharpness of the
  clusters partitioning. In particular, an higher cohesiveness among
  the judges within the first and second clusters with respect to
  those ones in the third cluster can be ascertained by the fact that
  a more coherent coloured pattern can be glimpsed from the beam of
  edges that originate from the first two clusters with respect to the
  last one, i.e. two different stacks can be distinguished, a red one
  and a blue one.}
\label{figure:judges}
\end{figure}

The second application is based on the dataset used in
\cite{doreian04} of the Supreme Court judges and their votes on a set
of issues. We have a {\em signed bipartite network} \cite{mrvar09}
with $N=9$ justices, $M=26$ issues and the expressed
votes\footnote{The table is filled with $+1$ if the judge voted in the
  majority for that issue and with $-1$ if he was in the minority in
  that decision. In case a $0$ is reported, it means that for that
  particular case the judge refrain to vote.}.\\ \indent In table
\ref{table:justice}, we present the result obtained with our algorithm
and the one obtained by Doreian \cite{doreian04}. At a first glance we
notice a remarkable similarity between the two results. However, it is
interesting to deepen the analysis by studying the data structure and
try to give a more detailed explanation for the differences. To this
end, we report in table (\ref{table:justiceData}) a permuted version
of the Supreme Court voting matrix, where the issues are blocked as in
\cite{Brusco06} and the judges are partitioned according to the
results from our algorithm, whereas in figure (\ref{figure:judges}) we
depict the bipartite network structure.  Looking at the first cluster,
containing {\it Scalia} and {\it Thomas}, and the second cluster,
composed by {\it Breyer}, {\it Ginsburg}, {\it Souter}, and {\it
  Stevens}, we can easily recognize a voting pattern remarkably
opposed one to each other and at the same time a coherent preference
expression within the groups.\\ \indent The unique puzzling doubt
concerns the allocation of {\it Rehnquist} in the group of {\it
  Kennedy} and {\it O'Connor} rather than in the group of {\it Scalia}
and {\it Thomas}. In order to further investigate this issue, we
decided to check the global likelihood value in the case where we move
{\it Rehnquist} in the first group. What we found is that the addition
of him to the group of {\it Scalia} and {\it Thomas} considerably
decreases the global likelihood.  This effect is a consequence of the
fact that our procedure recognizes the perfect dependence among these
last two actors, and therefore it prefers to allocate {\it Scalia }
and {\it Thomas} alone in one cluster in order to point out their
``positional equality'', and to group into the third cluster {\it
  O'Connor}, {\it Kennedy} and {\it Rehnquist}, which perfectly agree
over half of the issues. It is also worthwhile to remember that our
procedure is not allowed to give clusters of only one element.\\

\begin{table}[h]
\caption{Justice data}
\centering
\begin{tabular}{ccc|cccc}
\toprule
\multicolumn{3}{c}{Our method} & \multicolumn{4}{c}{Doreian \cite{doreian04}}\\
\midrule
Cluster 1 &Cluster 2 &Cluster 3 &Cluster 1 &Cluster 2 &Cluster 3 &Cluster 4 \\
\midrule
Scalia & Breyer & Kennedy & Scalia & Breyer & Kennedy & O'Connor \\
Thomas & Ginsburg  & O'Connor & Thomas & Ginsburg &     &            \\
       & Souter & Rehnquist  & Rehnquist  & Souter &   &            \\
       & Stevens &           &            & Stevens &  &           \\
\bottomrule
\end{tabular}
\label{table:justice}
\end{table}

\section{Conclusions}
\label{fine}

Clustering algorithms have increasingly assumed a central role for the
identification of communities in complex networks. In this paper, we
deal with a notion of community different from the classical
one: while the network clustering analysis, namely the
community detection, aims to identify clusters of densely connected
actors, we try to determine groups of actors that play a similar role
inside a certain organization basing on the characteristics or habits
that they exhibit.  In the social network literature, this is known as
positional analysis.\\ \indent To this end, we propose a new
clustering algorithm that can be applied to situations which are
suitably modelled through a {\em weighted bipartite network}.
Starting from the associated real-valued matrix, with the actors on
the rows, the features on the columns, and the weights as the
elements, we try to capture possible similarities among groups of
actors by analyzing the multivariate stochastic dependence among
them.\\ \indent The contribution of this paper has to be found in the
novelty of the methodological approach we propose for positional
analysis that is based on the detection of the intrinsic multivariate
stochastic dependence among groups of actors and in the development of
a related algorithm that uses copula functions in order to model these
dependence structures. Furthermore, this algorithm directly operates
on the matrix describing the actor-feature relationships, differently
from many other algorithms that collapse the information of the
bipartite network to a unipartite one and then apply the classical
clustering procedure. In fact, this kind of operation can cause a lost
of information and a consequent erroneous cluster identification.
Another advantage of our technique is that it finds the optimal
partition, without fixing {\it a priori} the number of clusters and/or
the number of elements per cluster (as, on the contrary, for other
clustering algorithms). Furthermore, our algorithm is able to work
directly on any matrix, binary or weighted with real
numbers.\\ \indent This is the first attempt of application of this
methodology to the network field and so it is not surprising that
there is still a issue to be addressed and that we leave for future
research.  Indeed, the major drawback of our algorithm concerns the
high computational burden that it bears, as a consequence of the fact
that it explores all the possible combinations of groups of
actors. Since our first purpose was to understand the potentiality of
such a new approach, we have not tried to develop any optimized
version of our algorithm yet.  However, we are convinced that a deeper
computational study of the behaviour of the algorithm will give some
insights on a possible criterion that could be exploited for a
reduction of the exploration procedure. From a first glance, our
suggestion to tackle this issue is to adopt an agglomerative or a
divisive approach as it is commonly used in the community detection
literature, based on some threshold on the log-likelihood function. In
such a case, the log-likelihood function could be computed only for
those groups of actors for which there is an interest with a
consequent reduction of the computational burden.  
\newpage

\noindent\textbf{Acknowledgment} \newline The authors acknowledge
support from CNR PNR Project ``CRISIS Lab'' and from the MIUR (FIRB
project RBFR12BA3Y). Moreover, Irene Crimaldi is a member of the
``Gruppo Nazionale per l'Analisi Matematica, la Pro\-ba\-bi\-li\-t\`a
e le loro Applicazioni (GNAMPA)'' of the ``Istituto Nazionale di Alta
Matematica (INdAM)''.\\ The authors thank Daniel B. Larremore for an
useful remark on the first version of this work.

\newpage

\appendix

\section*{Appendix}

\section{Tables}
 
\begin{table}[h]
	\caption{Supreme Court voting data.}
	\centering
	\begin{tabular}{ccccc|ccc|cc}
	\toprule
	Issue ($M=26$) & \multicolumn{9}{c}{Supreme Court justice ($N=9$)}\\
	\cmidrule{2-10}
	                           & Br &  Gi &  St &  So &  OC &  Ke &  Re &  Sc &  Th   \\
	\midrule
	President election         & -1  &  -1  &  -1  &  -1  &  1  &  1  &  1  &  1  &  1  \\
	Federalism                 & -1  &  -1  &  -1  &  -1  &  1  &  1  &  1  &  1  &  1  \\
	Clean Water                & -1  &  -1  &  -1  &  -1  &  1  &  1  &  1  &  1  &  1  \\
	Title VI Disabilities      & -1  &  -1  &  -1  &  -1  &  1  &  1  &  1  &  1  &  1  \\
	Tobacco Ads                & -1  &  -1  &  -1  &  -1  &  1  &  1  &  1  &  1  &  1  \\
	Labour rights              & -1  &  -1  &  -1  &  -1  &  1  &  1  &  1  &  1  &  1  \\
	Property Rights            & -1  &  -1  &  -1  &  -1  &  1  &  1  &  1  &  1  &  1  \\
	\cmidrule{2-10}
	Citizenship                & -1  &  -1  &  1   &  -1  &  -1  &  1   &  1  &  1  &  1  \\
	Free Speech                & 1   &  -1  &  -1  &  -1  &  1   &  1   &  1  &  1  &  1  \\
	Seat Belts                 & -1  &  -1  &  -1  &  1   &  -1  &  1   &  1  &  1  &  1  \\
	United Foods               & -1  &  -1  &  1   &  1   &  -1  &  1   &  1  &  1  &  1  \\
	NYT Copyright              & -1  &   1  &  -1  &  1   &  1   &  1   &  1  &  1  &  1  \\
	Cannabis for Health        & 0  &   1  &  1   &  1   &  1   &  1   &  1  &  1  &  1  \\
	Clean Air Act              & 1   &   1  &  1   &  1   &  1   &  1   &  1  &  1  &  1  \\
	PGA vs. Handicapped        & 1   &   1  &  1   &  1   &  1   &  1   &  1  & -1  & -1  \\
	Illegal Search 3           & 1   &   1  &  -1  &  1   &  -1  &  -1  & -1  &  1  &  1  \\
	\cmidrule{2-10}
	Illegal Search 1           & 1  &  1  &  1  &  1  &  1  &  1  & -1  & -1  & -1  \\
	Illegal Search 2           & 1  &  1  &  1  &  1  &  1  &  1  & -1  & -1  & -1  \\
	Stay of Execution          & 1  &  1  &  1  &  1  &  1  &  1  & -1  & -1  & -1  \\
	Privacy                    & 1  &  1  &  1  &  1  &  1  &  1  & -1  & -1  & -1  \\
	\cmidrule{2-10}
	Immigration Jurisdiction   & 1  &  1  &  1  &  1  & -1  &  1  & -1  & -1  & -1  \\
	Detaining Criminal Aliens  & 1  &  1  &  1  &  1  & -1  &  1  & -1  & -1  & -1  \\
	Legal Aid for the Poor     & 1  &  1  &  1  &  1  & -1  &  1  & -1  & -1  & -1  \\
	\cmidrule{2-10}
	Voting Rights              & 1  &  1  &  1  &  1  &  1  & -1  & -1  & -1  & -1  \\
	Deporting Criminal Aliens  & 1  &  1  &  1  &  1  &  1  & -1  & -1  & -1  & -1  \\
	Campaign Finance           & 1  &  1  &  1  &  1  &  1  & -1  & -1  & -1  & -1  \\
	\bottomrule
	\multicolumn{10}{l}{\footnotesize The data on the Supreme Court 
judges can be found in \cite{doreian04}. 
The blocks of the issues are based on \cite{Brusco06}.}\\
	\end{tabular}
	\label{table:justiceData}
\end{table}

\begin{landscape}
	\begin{table}[h]
		\caption{Selected sample of $21$ HS2 products traded by each country}
		\centering
		\begin{tabular}{c|cccc|cccc|cccc}
		\toprule
		{\bf HS2} & \multicolumn{12}{c}{{\bf Countries}} \\
		\cmidrule{2-13}
		                   & France  & Germany & U.S.A & Canada & Iran & Kuwait & Saudi Arabia & Qatar & Somalia & Zimbabwe & Liberia & Burundi \\
		\midrule
		84                 & 11.14   & 17.83   & 15.20 & 7.05   & 0.41 & 0.09   & 0.29         & 0.11  & 0.04    & 0.57     & 0.54    & 0.86  \\
		87                 & 9.74    & 17.86   & 8.22  & 12.07  & 0.21 & 0.06   & 0.17         & 0.01  & 0.00    & 0.15     & 0.06    & 0.69  \\
		88                 & 8.86    & 2.41    & 3.63  & 2.34   & 0.02 & 0.04   & 0.08         & 0.00  & 0.00    & 0.12     & 0.00    & 0.00  \\
		85                 & 7.60    & 9.85    & 10.38 & 3.34   & 0.22 & 0.05   & 0.21         & 0.03  & 0.02    & 0.36     & 2.27    & 0.21  \\
		30                 & 6.06    & 4.62    & 4.01  & 1.44   & 0.10 & 0.03   & 0.05         & 0.01  & 0.00    & 0.13     & 0.04    & 0.02  \\
		90                 & 2.97    & 4.34    & 5.96  & 1.29   & 0.05 & 0.01   & 0.03         & 0.01  & 0.00    & 0.02     & 0.04    & 0.10  \\
		\cmidrule{6-9}
		27                 & 4.64    & 3.08    & 9.02  & 26.22  & 85.69& 93.33  & 87.79        & 92.33 & 0.00    & 1.87     & 17.29   & 0.00  \\
		39                 & 3.70    & 4.84    & 4.36  & 2.82   & 2.02 & 1.19   & 3.84         & 2.09  & 0.52    & 0.20     & 0.09    & 0.03  \\
		29                 & 2.65    & 2.47    & 3.38  & 1.32   & 2.53 & 3.38   & 4.19         & 1.50  & 0.00    & 0.02     & 0.10    & 0.00  \\
		\cmidrule{2-5}  \cmidrule{6-9}
		31                 & 0.09    & 0.27    & 0.39  & 1.96   & 0.50 & 0.43   & 0.45         & 1.03  & 0.00    & 0.21     & 0.00    & 0.00  \\
		1                  & 0.43    & 0.12    & 0.08  & 0.33   & 0.03 & 0.01   & 0.03         & 0.01  & 80.23   & 0.04     & 0.00    & 0.05  \\
		9                  & 0.06    & 0.19    & 0.07  & 0.11   & 0.14 & 0.00   & 0.01         & 0.00  & 0.00    & 0.58     & 0.15    & 78.11 \\
		71                 & 1.38    & 1.56    & 3.68  & 5.31   & 0.34 & 0.03   & 0.20         & 0.15  & 0.01    & 19.09    & 2.10    & 0.29  \\
		40                 & 1.66    & 1.31    & 1.14  & 0.93   & 0.05 & 0.09   & 0.01         & 0.00  & 0.00    & 0.23     & 46.07   & 0.04  \\
		97                 & 0.45    & 0.07    & 0.36  & 0.05   & 0.01 & 0.00   & 0.00         & 0.02  & 0.00    & 0.11     & 0.07    & 0.00  \\
		96                 & 0.24    & 0.17    & 0.10  & 0.02   & 0.01 & 0.00   & 0.00         & 0.00  & 0.00    & 0.03     & 0.00    & 0.02  \\
		95                 & 0.26    & 0.39    & 0.34  & 0.23   & 0.00 & 0.00   & 0.00         & 0.00  & 0.00    & 0.03     & 0.00    & 0.00  \\
		93                 & 0.07    & 0.06    & 0.22  & 0.05   & 0.00 & 0.00   & 0.00         & 0.00  & 0.00    & 0.00     & 0.20    & 0.00  \\
		92                 & 0.04    & 0.05    & 0.05  & 0.02   & 0.00 & 0.00   & 0.00         & 0.00  & 0.00    & 0.00     & 0.00    & 0.02  \\
		91                 & 0.28    & 0.11    & 0.06  & 0.01   & 0.00 & 0.01   & 0.00         & 0.02  & 0.00    & 0.00     & 0.22    & 0.00  \\
		86                 & 0.20    & 0.36    & 0.23  & 0.10   & 0.00 & 0.00   & 0.00         & 0.00  & 0.00    & 0.02     & 0.01    & 0.00  \\
		\bottomrule
		\multicolumn{13}{l}{\footnotesize The table contains 
the percentage on the total amount of export for some product 
categories; while we applied the algorithm 
directly on the export values for all categories.}\\
		\end{tabular}
		\label{table:selection}
	\end{table}
\end{landscape}

\section{Archimedean family of copulas}
\label{copula}

Here we recall just the principal copula functions belonging to the
Archimedean family that we employ in our simulations and real data
analysis. \\ \indent The Archimedean family is defined using a
specific function $\phi$, known as the {\em generator}, by means of
the formula
\begin{equation*}
C({\bf u})=\phi^{-1}(\phi(u_1)+...+\phi(u_d)).
\end{equation*}
Different functional forms of the generator entail different
dependence structures. The principal Archimedean copulas are the
following.
\begin{itemize}
\item {\bf Gumbel copula}. The generator is given by
  $\phi(u)=(-\ln(u))^\theta$ and so the Gumbel copula is defined as
\begin{equation*}
C^{Gu}({\bf u};\theta)=
\exp \left\{-\left[\sum^d_{i=1}(-\ln u_i)^\theta \right]^\frac{1}{\theta} 
\right\},\qquad \theta\in\Theta=[1,+\infty).
\end{equation*}
The parameter $\theta$ tunes the degree of the dependence. In particular, 
the value $\theta=1$ corresponds to independence 
(indeed, we get $C^{Gu}({\bf u};1)=\prod_{i=1}^d u_i$).
\item {\bf Clayton copula}. The generator is given by
  $\phi(u)=(u^{-\theta}-1)/\theta$ and so the Clayton copula is defined as
\begin{equation*}
C^{Cl}({\bf u};\theta)= 
\left[ \sum^d_{i=1} u_i^{-\theta}-d+1\right]^{-\frac{1}{\theta}},
\qquad \theta\in\Theta=(0,+\infty).
\end{equation*}
\item {\bf Frank copula}. The generator is given by $\phi(u)=-\ln
  \left(\frac{\exp(-\theta u)-1}{\exp(-\theta)-1}\right)$ and so the
  Frank copula is defined as
\begin{equation*}
C^{Fr}({\bf u};\theta)=
-\frac{1}{\theta} 
\ln \left(1+ 
\frac{\prod^d_{i=1}(\exp(-\theta u_i)-1)}{(\exp(-\theta)-1)^{d-1}}\right),
\qquad \theta\in\Theta=(0,+\infty).
\end{equation*}
\end{itemize}
Also for these two last copulas, the parameter $\theta$ controls the
degree of the dependence.\\


\begin{thebibliography}{99}


\bibitem{aicher14} Aicher C, Jacobs AZ, Clauset A (2014) Learning
  Latent Block Structure in Weighted Networks. arXiv preprint
  arxiv:1404.0431v1.

\bibitem{asratian98} Asratian AS, Denley T, H\"{a}ggkvist R (1998)
  Bipartite Graphs and their Applications. Cambridge University Press.

\bibitem{barabasi02} Barab\'{a}si AL, Jeong H, N\'{e}da Z, Ravasz E,
  Schubert A, Vicsek T (2002) Evolution of the social network of
  scientific collaboration. Physica A 311: 590-614.

\bibitem{barber07} Barber MJ (2007) Modularity and community detection
  in bipartite networks. Phys Rev E Stat Nonlin Soft Matter Phys 76:
  066102.

\bibitem{borgatti97} Borgatti SP, Everett MG (1997) Network analysis
  of 2-mode data. Social Networks 19: 243-269.

\bibitem{borgatti08} Borgatti SP (2008) 2-Mode Concepts in Social
  Network Analysis. In Meyers RA. Encyclopedia of Complexity and
  System Science. Springer.

\bibitem{borgatti11} Borgatti SP, Halgin D (2011) Analyzing
  Affiliation Network. In Carrington P, Scott J (eds). The Sage
  Handbook of Social Network Analysis.

\bibitem{Brandes07} Brandes U, Lerner J (2007) Role equivalent Actors
  in Networks. In: Obiedkov S, Roth C. ICFCA Satellite Workshop on
  Social Network Analysis and Conceptual Structures: Exploring
  Opportunities.
  
\bibitem{Brusco06} Brusco M, Steinley D (2006) Inducing a blockmodel
  structure of two-mode binary data using seriation
  procedures. Journal of Mathematical Psychology 50: 468-477.

\bibitem{Brusco07} Brusco M, Steinley D (2007) A variable neighborhood
  search method for generalized blockmodeling of two-mode binary
  matrices. Journal of Mathematical Psychology 51: 325-338.

\bibitem{Brusco11} Brusco M (2011) Analysis of two-mode network data
  using nonnegative matrix factorization. Social Networks 33: 201-210.

\bibitem{cerina2014} Cerina F, Chessa A, Pammolli F, Riccaboni M
  (2014) Network communities within and across borders. Scientific
  Reports 4: 1-7.

\bibitem{Good10} Good BH, de Montjoye YA, Clauset A (2010) The
  performance of modularity maximization in practical contexts. Phys
  Rev E Stat Nonlin Soft Matter Phys 81: 046106.

\bibitem{Davis97} Davis GF, Greve HR (1997) Corporate elite networks
  and governance changes in the 1980. American journal of sociology
  103(1): 1-37.  1997.

\bibitem{dhillon01} Dhillon IS (2001) Co-clustering documents and
  works using bipartite spectral graph partitioning.  Proceedings of
  the seventh international conference on knowledge discovery and data
  mining: 269-274.

\bibitem{diLascio12a} Di Lascio FM, Giannerini S (2012) A copula-based
  Algorithm for Discovering Patterns of Dependent
  Observations. Journal of Classification 29(1): 50-75.

\bibitem{doreian04} Doreian P, Batagelj V, Ferligoj A (2004)
  Generalized blockmodeling of two-mode network data. Social Networks
  26: 29-53.

\bibitem{everett13} Everett MG, Borgatti SP (2013) The dual-projection
  approach for two-mode networks. Social Networks 35: 204-210.

\bibitem{fortunato10} Fortunato S (2010) Community detections in
  graphs. Physics Reports 486: 75-174.

\bibitem{Girvan02} Girvan M, Newman MEJ (2002) Community structure in
  social and biological networks. Proc Natl Acad Sci U S A 99(12):
  7821-7826.

\bibitem{Guimera05} Guimer\`a R, Uzzi B, Spiro J, Amaral LAN (2005) Team
  assembly mechanisms determine collaboration network structure and
  team performance. Science 308: 697-702.

\bibitem{Guimera07} Guimer\`a R, Sales-Pardo M, Amaral LAN (2007) Module
  identification in bipartite and directed networks. Phys Rev E Stat
  Nonlin Soft Matter Phys 76: 036102.

\bibitem{hoeffding40} Hoeffding W (1994) Scale invariant correlation
  theory. In Fisher NI, Sen PK (eds). The Collected Works of Wassily
  Hoeffding. Springer Series in Statistics. pp. 57-104.

\bibitem{hoppner99}
Hoppner F, Klawonn F, Kruse R, Runkler (1999) Fuzzy cluster analysis.
John Wiley and Sons, Ltd.

\bibitem{joe97} Joe H (1997) Multivariate Models and Dependence
  Concepts. Chapman and Hall.

\bibitem{larremore13} Larremore DB, Clauset A, Buckee CO (2013) A
  Network Approach to Analyzing Highly Recombinant Malaria Parasite
  Genes. PLoS Comput Biol 9(10):
  e1003268. doi:10.1371/journal.pcbi.1003268

\bibitem{larremore14} Larremore DB, Clauset A, Jacobs AZ (2014)
  Efficiently inferring community structure in bipartite
  networks. arXiv:1403.2933.

\bibitem{lohr12} Lohr S (2012) The age of big data. New York Times
  [Internet]. Available from:
  http://wolfweb.unr.edu/homepage/ania/NYTFeb12.pdf.

\bibitem{Mariolis75} Mariolis P (1975) Interlocking directorates and
  control of corporations. Social Science Quarterly 56(3): 425-439.

\bibitem{mrvar09} Mrvar A, Doreian P (2009) Partitioning Signed
  Two-Mode Networks. Journal of Mathematical Sociology 33: 196-221.

\bibitem{nelsen06} Nelsen RB (2006) An introduction to
  Copulas. Springer Series in Statistics.

\bibitem{newman01} Newman MEJ (2001) Scientific collaboration
  networks. Network construction and fundamental results. Phys Rev E
  Stat Nonlin Soft Matter Phys 64: 016131.  2001.

\bibitem{sklar59} Sklar A (1959) Fonctions de la repartition a n
  dimensions et leurs marges.  Publications de l'Institute de
  Statistique de l'Universite de Paris 8: 229-231.

\bibitem{trivedi} Trivedi PK, Zimmer DM (2005) Copula modeling: an
  introduction for practitioners. Foundations and trends in
  Econometrics 1(1): 1-111.

\bibitem{tzekina08} Tzekina I, Danthi K, Rockmore DN (2008) Evolution
  of community structure in the world trade web. European Physics
  Journal B 63: 541-545.

\bibitem{wasser94} Wasserman F, Faust K (1994) Social Network
  Analysis: Methods and Applications. Cambridge University Press.

\bibitem{Zhou07} Zhou T, Ren J, Medo M, Zhang Y (2007) Bipartite
  network projection and personal recommendation. Phys Rev E Stat
  Nonlin Soft Matter Phys 76: 046115.

\bibitem{zib07} Ziberna A (2007) Generalized blockmodeling of valued
  networks.  Social Networks 29: 105–126.
\end{thebibliography}
\end{document}